\documentclass[11pt,twoside]{article}

\usepackage{asp2006}
\usepackage{epsf}
\usepackage{lscape}

\begin{document}
\title{Magnetic fields, winds and X-rays of massive stars: A spectropolarimetric survey of the Orion cluster}   

\author{ V. Petit$^1$, G. Wade$^2$, L.Drissen$^1$, T. Montmerle$^3$, E. Alecian$^2$ }   

\affil{ $^1$ D\'ept. de physique, g\'enie physique et optique, Centre de recherche en astrophysique du Qu\'ebec, Universit\'e Laval, Qu\'ebec, Canada
\\ $^2$ Dept. of Physics, Royal Military College of Canada, Kingston, Canada
\\$^3$ Laboratoire d'Astrophysique de Grenoble, Universit\'e Joseph Fourier, Grenoble Cedex, France}   

\begin{abstract} 

In massive stars, magnetic fields are thought to confine the outflowing radiatively-driven wind, resulting in X-ray emission that is harder, more variable and more efficient than that produced by instability-generated shocks in non-magnetic winds. Although magnetic confinement of stellar winds has been shown to strongly modify the mass-loss and X-ray characteristics of massive OB stars, we lack a detailed understanding of the complex processes responsible. The aim of this study is to examine the relationship between magnetism, stellar winds and X-ray emission of OB stars. In conjunction with a Chandra survey of the Orion Nebula Cluster, we carried out spectropolarimatric ESPaDOnS observations to determine the magnetic properties of massive OB stars of this cluster.
We found of two new massive magnetic stars in the Orion Nebula Cluster: HD\,36982 and HD\,37061, for which the estimated dipole polar strengths are $1150^{+320}_{-200}$\,G and $620^{+220}_{-170}$\,G, respectively.
However, the apparent lack of clear correlation between X-ray indicator and the presence of a magnetic fields brings forth new challenges for understanding the processes leading to X-ray emission in massive stars.

\end{abstract}

\section{Introduction}

Magnetic fields are well known to produce X-rays in late-type convective stars like the Sun.
On the other hand, X-rays emission from massive stars is thought to have a different origin. Their powerfull winds are driven by radiation, which is an unstable process. These instabilities result in collisions between wind streams of different velocity, resulting in small shocks that generate X-ray emission \citep{1999ApJ...520..833O,1980ApJ...241..300L}.
Typically, this emission is soft and stable. Also, empirical studies have shown that the X-ray luminosity will be about $10^{-7}$ times the bolometric luminosity for O stars, and will decrease rapidly between spectral type B1 and B3 \citep{1997ApJ...487..867C}.

However, some observations are not consistent with such a model. 
The best exemple is $\theta^1$\,Ori\,C, in the Orion Nebula Cluster. This young O star displays hard and efficient X-rays, and shows a modulated X-ray light curve \citep{1997ApJ...478L..87G}, which varies with the same period as lines in the UV and optical spectrum \citep{1993A&A...274L..29S,1996A&A...312..539S,1994ApJ...425L..29W}.

\citet{1997ApJ...485L..29B} explained this peculiar X-ray emission in terms of the ``magnetically confined wind shock'' model (MWCS). In this model, the dipolar stellar magnetic field channels the outflowing wind along the field lines, resulting in a closed magnetosphere with large-scale equatorial shocks which enhance the X-ray emission and possibly modulate it with stellar rotation. Such a magnetic field ($1.1\pm0.1$\,kG) was subsequently discovered by \citet{2002MNRAS.333...55D}.

From that exemple, we can speculate that X-ray enhancement and periodic modulation may be associated with the presence of a magnetic field, in the form of a magnetospheric confining of the wind.

\subsection{The Orion Nebula cluster}

\begin{figure}
	\plotone{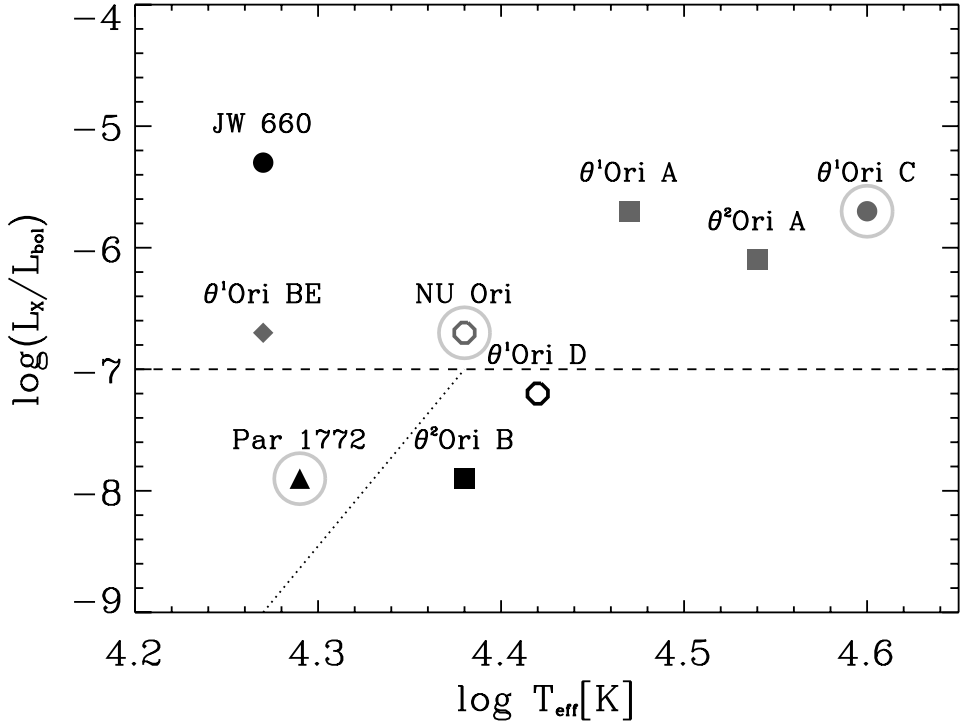}
	\caption{\label{fig1}X-ray efficiency of the ONC massive stars as a function of effective temperature (Stelzer et al. 2005). The 3 detected stars are circled. Filled symbols are for stars with indirect indications of the presence of a magnetic field, and grey symbols are for confirmed or suspected binaries. Plotting symbols indicate the following properties: circles are for stars showing possible X-ray rotational modulation, squares are for T\,Tauri type X-ray emission, triangles are chemically peculiar (CP) stars, and the diamond star was not observed. The dotted lines indicates the typical efficiency for massive O-type stars, and a qualitative illustration of the sharp decrease in the B-type stars range.}
\end{figure}

The Orion Nebula Cluster (ONC) presents a unique opportinuty to characterize the magnetic fields of nearby massive OB stars.
A \textit{Chandra} large program was dedicated to observe the ONC in X-rays: The \textit{Chandra} Orion Ultradeep Project (COUP) consisted of a 10-day exposure of the ONC \citep{2005ApJS..160..319G}.
\citet{2005ApJS..160..557S} have studied the COUP massive star sample, containing 9 massive OB stars for which radiatively-driven winds should be the dominant mechanism for X-ray emission. The list of stars, along with their properties, can be found in Table \ref{tab1}.
\citeauthor{2005ApJS..160..557S} found a large variety of X-ray properties for the ONC massive stars. For example there is a large scatter (3 orders of magnitude) around the empirical relation between the X-ray luminosity and the bolometric luminosity ($L_\mathrm{X} / L_\mathrm{bol} = 10^{-7}$), and also a large incidence of variability and flares. The X-ray properties are summarised in Figure \ref{fig1}.
Altogether, the X-ray results, combined with other indicators, strongly suggest that all but two stars (NU\,Ori and $\theta^1$\,Ori\,D) of the ONC are magnetic \citep{2005ApJS..160..557S}.

\section{Observations}

\begin{table}[t]
	\caption{\label{tab1}Target properties and longitudinal field error bar achieved}
	\smallskip
	\begin{center}
		{\small
		\begin{tabular}{cccccc}   
			\tableline
			\noalign{\smallskip}
		 		& & V  & Type & vsin$i$  & $\sigma_B$ \\
				& &      &           & [km/s]                & [G] \\
			\noalign{\smallskip}
			\tableline
			\noalign{\smallskip}
				$\theta^{1}$\,Ori\,C & HD\,37022	& 5.1 & O7      	& 24     & 35\\
				$\theta^{2}$\,Ori\,A & HD\,37041    	& 5.1 & O9.5   	& 131   & 42\\
				$\theta^{1}$\,Ori\,A & HD\,37020    	& 6.7 & B0   	& 55   & 18\\
				$\theta^{1}$\,Ori\,D & HD\,37023   	& 6.7 & B0.5	& 49     & 22\\
				NU\,Ori 		       & HD\,37061  	& 6.8 & B1 	& 180   & 54\\
				$\theta^{2}$\,Ori\,B & HD\,37042   	& 6.0 & B1 	& 32    & 8\\
				Par\,1772    	       & HD\,36982	& 8.4 & B2       	& 98  &  61\\
				JW\,660		       & 		 	& 9.7 & B3       	& 210  &  258\\
				$\theta^1$\,Ori\,BE &HD\,37021  & 8.0 & B3	& 220 & $-$ \\
			\noalign{\smallskip}
			\tableline
		\end{tabular}
		}
	\end{center}
\end{table}

To examine the effect of magnetic fields on the winds of the massive stars in the ONC, we conducted spectropolarimetric observations in order to directly detect and characterise potential magnetic fields in these stars, through the circular polarisation induced by the Zeeman effect. 
We used the ESPaDOnS spectropolarimater at CFHT in January 2006 and March 2007 to observe 8 of the 9 ONC massive stars. Additional measurements of $\theta^1$\,Ori\,C and Par\,1772 were taken with ESPaDOnS in December 2007 and with ESPaDOnS's twin Narval, installed at T\'elescope Bernard Lyot, in November 2007. 

ESPaDOnS and Narval are ideal instruments for this study. 
The amplitude of the circular polarisation signature is dependent on the broadening of the line, either intrinsic or rotational. The stars in the sample have high projected rotational velocity, hence the need for a high signal-to-noise ratio. 
Furthermore, the number of spectral lines in the visible range is smaller for early-type stars than in late-type stars, so the gain in s/n achievable from spectral line combination methods, such as the Least-Squares Deconvolution procedure (LSD) developed by \citet{1997MNRAS.291..658D}, is reduced. 
With the large telescope aperture, we can achieve the desired s/n in an amount of time that will not cover a too large fraction of the rotation period. A long exposure risks smearing the variable polarisation signature as the star rotates. 
These echelle spectrographs have a large spectral coverage enabling the use of as many spectral lines as possible.
Finally, with the high resolving power ($R=65,000$), we measure the circular polarisation within resolved spectral lines. 
Only a small amount of stellar rotation will induce doppler shifts across the stellar disk, that will separate opposite polarisation components that would otherwise cancel out. Therefore we are sensitive to the presence of a magnetic field even when the the mean longitudinal component is null (e.g. dipolar field at ``crossover'' phases).

A complete circular polarisation observation consists of series of four sub-exposures between which the polarimeter quarter-wave plate is rotated back and forth between position angles, which makes it possible to reduce systematic errors. For a complete description of the observation and reduction procedure with the \textsc{libre-esprit} package provided by CFHT, see \citet{1997MNRAS.291..658D}.

From our observations, we found clear Stokes V signatures for three stars: the previously-detected $\theta^1$\,Ori\,C, as well as HD\,36982 (=Par\,1772, LP\,Ori) and HD\,37061 (=NU\,Ori) \citep{2008MNRAS.387L..23P}. The magnetic stars are encircled in Figure \ref{fig1}. The last column of Table \ref{tab1} gives the error bar on the longitudinal field derived by integrating over the line profile, as described by \citet{1997MNRAS.291..658D}. 
The Stokes V signatures were directly modelled with the polarised LTE radiative transfer code \textsc{zeeman2} \citep{1988ApJ...326..967L,2001A&A...374..265W}, leading to an inferred surface dipolar field strength of $1150^{+320}_{-200}$\,G and $620^{+220}_{-170}$\,G for HD\,39682 and HD\,37061 respectively. 

\section{Discussion}

The role of magnetic fields in X-ray production remains poorly understood. As shown in Figure \ref{fig1}, X-ray variability and enhancement is not necessarily correlated with the presence of a magnetic field. Furthermore, magnetic fields are observed even in the absence of X-ray indications. 
Theoretical work, such as numerical MHD simulations of magnetically confined stellar winds done by \citet{2008MNRAS.385...97U} will help disentangle various effect such as rotation, geometry of the magnetic fields, etc. that could explain the current observations. However, we also need more observations to constrain those simulations. A large CFHT observing program was allocated to the Magnetism in Massive Stars (MiMeS) collaboration that will allow us to better understand the interaction of stellar winds and magnetic fields.

\acknowledgements 
VP acknowledges support from Fonds qu\'eb\'ecois de la recherche sur la nature et les technologies. 
GAW acknowledges support from the Discovery Grants programme of the Natural Science and Engineering Research Council of Canada. 
LD acknowledges support from the Canada Research Chair programme and the Discovery Grants programme of the Natural Science and Engineering Research Council of Canada. 
EA is supported by the Marie Curie FP6 program.

\end{document}